\documentstyle[aps,prc,epsfig]{revtex}
\begin{document}
\newcommand{\beq}{\begin{equation}}
\newcommand{\eeq}{\end{equation}}
\newcommand{\bea}{\begin{eqnarray}}
\newcommand{\eea}{\end{eqnarray}}
\newcommand{\bfig}{\begin{figure}}
\newcommand{\efig}{\end{figure}}
\newcommand{\ie}{{\it i.e.}}
\newcommand{\bce}{\begin{center}}
\newcommand{\ece}{\end{center}}
\newcommand{\eg}{{\it e.g.}}
\newcommand{\Eg}{{\it E.g.}}
\newcommand{\etal}{{\it et al.}}
\def\lsim{\mathrel{\rlap{\lower4pt\hbox{\hskip1pt$\sim$}}
    \raise1pt\hbox{$<$}}}	  
\def\gsim{\mathrel{\rlap{\lower4pt\hbox{\hskip1pt$\sim$}}
    \raise1pt\hbox{$>$}}}	  
\title{Dileptons in High-Energy Heavy-Ion Collisions}

\author{Ralf Rapp}
\address{Department of Physics and Astronomy, SUNY Stony Brook,  
New York 11794-3800, U.S.A.}
\maketitle


\begin{abstract}
The current status of our understanding of 
dilepton production in ultrarelativistic heavy-ion collisions is 
discussed with special emphasis on signals from the (approach towards) 
chirally restored and deconfined phases. In particular, recent results
of the CERN-SPS low-energy runs are compared to model predictions
and interpreted. Prospects for RHIC experiments are given.
\end{abstract}


\section{Introduction}
The rich physics potential of electromagnetic observables (photons 
and dileptons) has been clearly demonstrated over the past decade  
of heavy-ion experiments, especially at the CERN-SPS.
Although the final spectra constitute a superposition of emission
from the entire space-time history of a nucleus-nucleus collision, 
the original idea~\cite{Shu80}
 of identifying signals from (thermalized?!) hot and 
dense phases of strongly interacting matter could be realized.  
Indeed, an excess of radiation in central Pb(158AGeV)-Au and 
Pb(158AGeV)-Pb collisions 
as compared to baseline $p$-$p$ and $p$-$A$ measurements 
has consistently been observed, most prominently for electron pairs 
at low invariant mass ($M\le 1$~GeV) by CERES/NA45~\cite{na45-158}, 
muon pairs at intermediate mass (1.5~GeV~$\le M \le$~3~GeV) by 
NA50~\cite{na50-int} and photons at transverse momenta $q_t\gsim1.5$~GeV 
by WA98~\cite{wa98}.  
>From the theoretical side it should be emphasized that both dilepton and 
photon emission rates of a thermal medium  
are based on the same quantity, \ie, the (imaginary part of the) 
correlation function of the electromagnetic (e.m.) current (or photon 
selfenergy), Im~$\Pi_{\rm em}$. Thus a consistent description 
of both photon and dilepton observables within a common approach  
(coupled with a realistic space-time evolution for a heavy-ion 
collision) is mandatory (see refs.~\cite{RW00,PT02} for recent reviews). 
Whereas real photons are related to   
the $M^2\to 0^+$ limit of $\Pi_{\rm em}$\footnote{The 
e.m. correlator is also directly relevant to studies 
of charge fluctuations. The fluctuation content of a locally thermalized 
system is directly proportional to the static spacelike limit 
of $\Pi_{\rm em}$.}, the production of  
timelike photons (=dileptons) requires $M^2>0$. The latter therefore 
carry additional dynamical information as, \eg, encoded in the 
low-lying (nonperturbative) vector-meson ($J^P=1^-$) resonance 
excitations ($\rho$, $\omega$ and $\phi$) of the QCD vacuum.
This renders the low-mass region an  ideal regime to study 
pertinent in-medium effects in connection with (the approach towards)
chiral symmetry restoration, as will be further detailed below.
At invariant masses beyond 1.5~GeV, the e.m. correlator becomes 
perturbative in nature, characterized by a rather structureless 
continuum with a strength governed by free $q\bar q$ 
with controllable corrections in the strong coupling $\alpha_s$, 
temperature and density. With an approximately known emissivity 
the virtue of the probe resides in the magnitude of the signal 
as an indicator of the temperatures reached in the early stages. 
The same idea is the main motivation~\cite{Alam} behind direct photon
measurements. There is, however, a subtle but important difference:
the leading contribution to photon production is of order 
${\cal O}(\alpha \alpha_s)$ ($\alpha=1/137$: e.m. coupling constant), 
whereas the one for (intermediate-mass) dileptons is ${\cal O}(\alpha^2)$
and therefore under better theoretical control (albeit 
experimentally suppressed).

\section{Electromagnetic Radiation: Hadron Gas vs. QGP }
\subsection{Production Rate and E.M. Correlator in Free Space}
To leading order in $\alpha$, the emission rate of lepton pairs 
$l^+l^-$ ($l=e,\mu$) from thermally equilibrated matter per unit 
4-volume and 4-momentum reads~\cite{MT85} 
\beq
\frac{dR_{l^+l^-}^{therm}}{d^4q} = -\frac{\alpha^2}{\pi^3 M^2} 
 \ f^{Bose}(q_0;T) \ {\rm Im} \Pi_{\rm em}(M,q;\mu_B;T)
\label{rate}
\eeq
($M^2=q_0^2-q^2$).
The theoretical objective is to calculate the (thermal expectation 
value of the) correlation function of two e.m. currents,  
\beq
\Pi_{\rm em}^{\mu\nu}(q_0,q)=\int d^4y \ {\rm e}^{i q\cdot y} \  
 \langle 0 |j_{\rm em}^\mu(y) j_{\rm em}^\nu(0)|0 \rangle \ ,  
\eeq  
which, in principle, contains all orders in the strong interaction.  
In practice, one has to invoke approximations starting from either 
quark or hadronic degrees of freedom. Accordingly, the 
current operator takes the form 
\beq
j^\mu_{\rm em}= \left\{
\begin{array}{ll}
\frac{1}{\sqrt 2}j_\rho^\mu +\frac{\sqrt{2}}{6} j_\omega^\mu 
-\frac{1}{3} j_\phi^\mu \qquad \qquad \qquad \qquad
{\rm hadronic \  basis}
\vspace{0.3cm}
\\
\sum\limits_q e_q \bar q \gamma^\mu q = \frac{2}{3} \bar u \gamma^\mu u
-\frac{1}{3} \bar d \gamma^\mu d -\frac{1}{3} \bar s \gamma^\mu s 
\qquad {\rm quark \ basis} \ .
\end{array} \right.
\label{jem}
\eeq
In free space, the 'transition'  can be inferred from the famous 
$e^+e^- \to hadrons$ cross section: at low invariant mass,
$\sqrt s \equiv M \lsim 1.2$~GeV, it is  
saturated by $\rho$, $\omega$ and $\phi$ mesons  
(Vector Dominance Model = VDM), whereas beyond $M\simeq 1.5$~GeV, the 
cross section is determined by weakly correlated 
$q \bar q$ pairs with almost no impact from subsequent 
hadronization (until additional flavor thresholds are
reached where nonperturbative bound state formation occurs).     
Thus, the e.m. correlation function may be decomposed as
\beq
{\rm Im} \Pi_{\rm em}(s) = \left\{
\begin{array}{ll}
 \sum\limits_{V=\rho,\omega,\phi} \left(\frac{m_V^2}{g_V}\right)^2 \
{\rm Im} D_V(s) & , \ s \le s_{dual}
\vspace{0.3cm}
\\
-\frac{s}{12\pi} \ (1+\frac{\alpha_s(s)}{\pi} +\dots)  \ N_c 
\sum\limits_{q=u,d,s} (e_q)^2  & , \ s \ge s_{dual} \ .
\end{array}  \right.
\label{Piem}
\eeq

\begin{figure}[h]
\bce
\epsfig{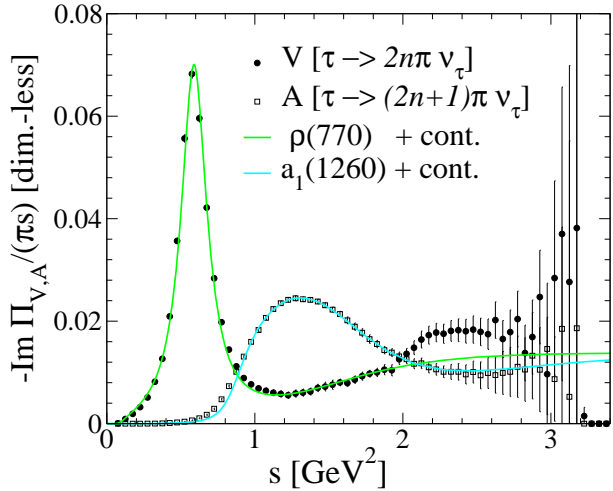}
\ece
\vspace{0.1cm}
\caption{ALEPH data for axial-/vector correlators as 
extracted from $\tau$ decays~\protect\cite{aleph}.  The curves 
are model calculations using $\rho$ and $a_1$ resonances 
plus perturbative continua.}
\label{fig_VAvac}
\end{figure}

For the investigation of medium effects through dilepton production,
an important question is how the spontaneous breaking of chiral symmetry 
(SBCS) manifests itself in the e.m. (or vector) correlator. A direct 
connection can be established in the isovector ($I=1$) channel: at
low mass, the pertinent hadronic current is entirely governed by the 
$\rho(770)$-meson which has a well-defined 'chiral partner' in form 
of the $a_1(1260)$-meson, which itself saturates the low-mass part of 
the ($I=1$) axialvector current. Recent data from 
$\tau$-decays by the ALEPH collaboration~\cite{aleph} 
nicely exhibit this feature, cf.~Fig.~\ref{fig_VAvac}.
For $s\le s_{dual}$, the isovector hadronic (axial-) vector correlator
 can therefore be written as
\beq
{\rm Im} \Pi_{V,A}^{I=1}(s)   = \left\{
\begin{array}{ll}
 \left(\frac{m_\rho^2}{g_\rho}\right)^2 \
{\rm Im} D_\rho(s) & , IJ^P=11^- 
\vspace{0.3cm}
\\
 \left(\frac{m_{a_1}^2}{g_{a_1}}\right)^2 \
{\rm Im} D_{a_1}(s)  -s f_\pi^2 \pi \delta(s-m_\pi^2) 
  & , IJ^P=11^+ \ .
\end{array}  \right.
\label{PiVAhad} 
\eeq 
The difference between the two is the manifestation of SBCS 
in the QCD vacuum. 
Note the appearance of the pion pole contribution (the Goldstone boson
of the symmetry breaking) with a residue given by the pion decay constant.
At higher masses  both correlators merge
into a continuum value determined by perturbation theory:
\beq
{\rm Im} \Pi_{V,A}^{I=1} =-\frac{s}{12\pi} \ 
    (1+\frac{\alpha_s(s)}{\pi} +\dots)  \ N_c \ \frac{1}{2} \ , 
\label{PiVApert}
\eeq
\ie, beyond $s\simeq s_{dual}$ effects due to SBCS no longer play a role. 

Chiral symmetry breaking can be further quantified in terms of the 
Weinberg sum rules~\cite{Wei67}, which are energy-weighted moments 
of the difference between vector and axialvector correlators, 
\eg,
\beq
f_\pi^2=- \int \frac{ds}{\pi s} ({\rm Im}\Pi_\rho -{\rm Im}\Pi_{a_1}) \ . 
\eeq
Kapusta and Shuryak~\cite{KS94}  have shown that these also hold in hot 
and dense matter which provides useful constraints on hadronic models. 
Chiral restoration obviously requires the two correlators to 
degenerate over the entire mass range, 
\ie, ${\rm Im} \Pi_\rho(s)={\rm Im} \Pi_{a_1}(s)$. 
How this is realized constitutes one of the main questions in the 
context of low-mass dilepton observables and related theoretical analyses.  
In this respect it is fortunate that the isovector channel dominates
the strength in the e.m. correlator;  for two flavors, \eg, the 
coefficient $\frac{1}{2}$ in eq.~(\ref{PiVApert}) amounts to $\sim$90\% 
of the total emissivity (coefficient $\frac{5}{9}$ in eq.~(\ref{Piem})).

\subsection{Medium Effects}
Following the decomposition suggested in the previous section,  
a schematic overview of in-medium effects is given in Tab.~\ref{tab_med}.

\begin{figure}[!t]
\bce
\epsfig{file=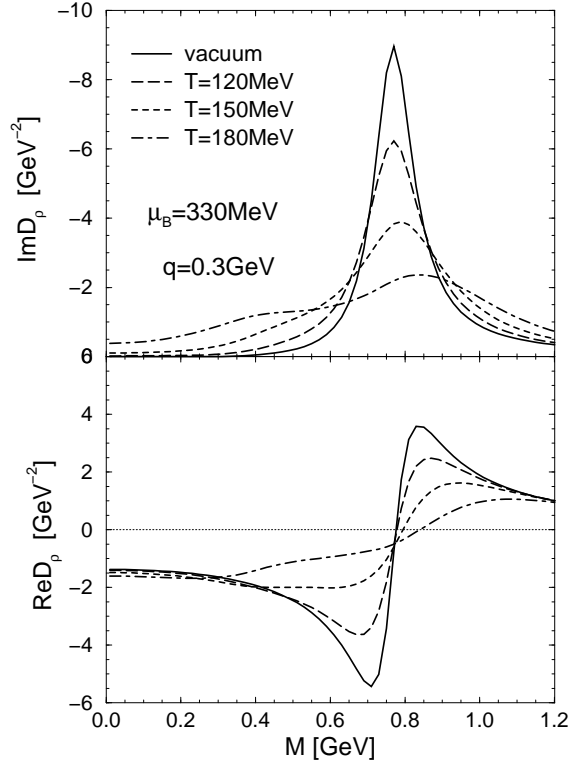,width=8cm}
\ece
\caption{Real and imaginary part of the $\rho$-meson 
propagator~\protect\cite{RW99} at finite temperatures and baryon 
densities.} 
\label{fig_drho}
\end{figure}

As emphasized before, the low-mass radiation from hadronic matter is
well suited to study in-medium modifications of light vector-mesons.
Their vacuum properties are intimately related to the spontaneous chiral
symmetry breaking; \eg, $M_V\simeq 2 M_q$, with the constituent quark
mass $M_q=400-500$~MeV driven by the $\langle\bar qq\rangle$-condensate.
Thus, even below $T_c$,  medium modifications of vector mesons ought to be
considered as precursor phenomena of chiral restoration.
Indeed, for low-mass lepton-pair emission the overall thermal
Bose factor in eq.~(\ref{rate}) induces a relatively moderate
dependence on temperature, which is
largely compensated by the volume increase
of the expanding and cooling fireball in a heavy-ion collision.
Thus, one is rather sensitive to medium effects in the spectral
distributions themselves. These have been extensively studied in the
literature.
Starting point are typically effective, chirally invariant
lagrangians for hadronic interaction vertices with coupling constants
and formfactors determined by experiment. Based on nuclear and
finite-temperature many-body techniques one evaluates a vector-meson
selfenergy
$\Sigma_\rho= \Sigma_{\rho\pi\pi} + \Sigma_{\rho M} + \Sigma_{\rho B}$
(similar for $\omega$ and $\phi$) arising
from interactions with the surrounding
hot ($M=\pi, K, \rho, \dots$ ) and dense ($B= N, \Lambda, \Delta,
\dots$) hadron gas (HG); $\Sigma_{\rho\pi\pi}$ incorporates medium
modifications of the free $\rho\to\pi\pi$ decay.
The generic outcome of such calculations is a strong increase of
$|{\rm Im} \Sigma_\rho|$ with temperature and density, which broadens
the $\rho$-meson spectral function, Im~$D_\rho$, beyond recognition
of any resonance structure, cf. Fig.~\ref{fig_drho}.
At comparable densities, nuclear effects prevail over the ones
induced by thermal pions (roughly speaking, interactions
with pions are 'Goldstone'-protected).
\begin{table}[!h]
\caption{Summary of medium effects in thermal dilepton production.}
\vspace{0.2cm}
\begin{tabular}{c|c|c}
          & Hadron Gas  & Quark-Gluon Plasma \\ 
\hline
               & \underline{in-medium $\rho$, $\omega$, $\phi$:}  & 
                 \underline{perturbative QCD:}         \\
  Low          &  effective chiral Lagrangian  + VDM & 
              HTL-resummed $q \bar q$ annihilation~\protect\cite{BPY90} \\
  Mass      &  + finite-$T$ / -$\mu_B$ field theory~\protect\cite{RW00}    &
             + LPM effect~\protect\cite{CGR93} \\
$M\lsim 1$~GeV &$D_V=[M^2-m_V^2-\Sigma_V(M,q;\mu_B,T)]^{-1}$    &  
                                     \underline{non-pert. QCD:}  \\ 
        &   & gluon condensates ($T\gsim T_c$)~\protect\cite{LWZH99}\\   
\hline
Intermediate   &  $\pi a_1 \to l^+l^-$ annihilation~\protect\cite{LG98}  &   
                      'bare' $\alpha_s$ corrections  \\
Mass           &  $\hat{=}$ chiral V-A mixing~\protect\cite{DEI90}        & 
                  to $q\bar q$ annihilation~\protect\cite{AR89,KW00}  \\
$M\gsim 1$~GeV &$\Pi_{V,A}=(1-\epsilon)~\Pi_{V,A}^\circ 
                       +\epsilon~\Pi_{A,V}^\circ$
                 &  order ${\cal O}(\alpha_s\frac{T^2}{M^2})$    
\end{tabular}
\label{tab_med}
\end{table}

In the invariant-mass window between 1 and 1.5~GeV, the main source
of hadronic dilepton production is due to $\pi a_1$ 
annihilation~\cite{LG98}. This, in fact, can be understood as a 
pion-induced mixing between the axialvector and vector correlator 
(and vice versa) to lowest order in temperature, characterized by the 
parameter $\epsilon=T^2/6f_\pi^2$ (chiral limit)~\cite{DEI90}.
When extrapolated to high temperatures, this 'chiral mixing' leads to 
a mutual degeneracy of $V$- and $A$-correlators not too far from the 
critical temperature as extracted from lattice gauge theory, 
$T_c=170$-190~MeV.     

Another line of treating medium effects employs chiral lagrangians
in mean-field approximation coupled with density-dependent corrections 
to masses and coupling constants. \Eg, within the 
so-called Brown-Rho scaling scenario~\cite{BR01}, all hadron masses 
(except for Goldstone bosons) 'drop' with increasing temperature and 
density.

Turning to the QGP phase, it turns out that, at low mass, the correct 
leading-order in $\alpha_s$ result requires a resummation of thermal 
propagators accompanied by vertex corrections, which is achieved
within the so-called Hard Thermal Loop (HTL) framework~\cite{BPY90}. 
Although the strong coupling constant $g_s$ is not really small 
for practical applications in the QGP phase, $T\simeq$~(1-3)~$T_c$,   
the main qualitative result of a substantial enhancement of the rate
towards small $M$ (due to 'Bremsstrahlung'-type processes) should 
be rather robust. At higher masses, $M\gg T$, naive perturbation 
theory becomes applicable again, with moderate 
corrections~\cite{AR89,KW00}. 

\begin{figure}[!ht]
\vspace{-0.5cm}
\bce
\epsfig{file=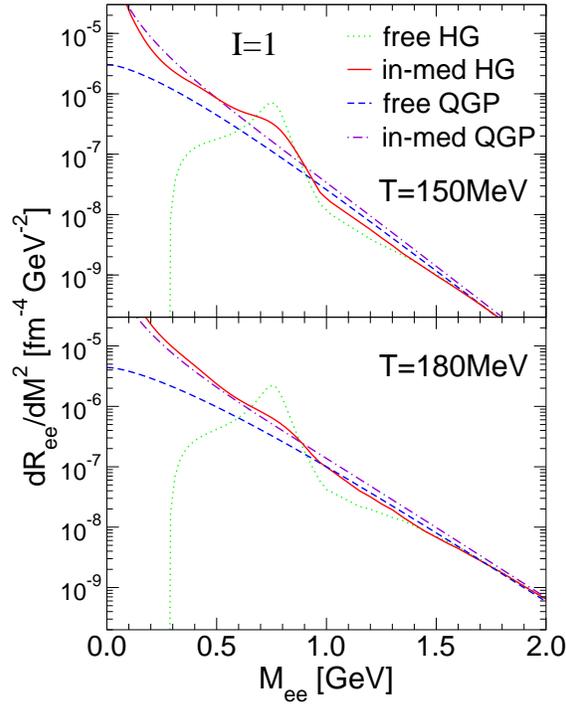,width=8cm}
\ece
\caption{Equilibrium dilepton production rates from hadronic and QGP
matter.}
\label{fig_rate}
\end{figure}

Fig.~\ref{fig_rate} summarizes the various features graphically
in terms of the 3-momentum integrated production rate
\beq
\frac{dR_{ee}}{dM^2} = -\frac{\alpha^2}{\pi^3 M^2}
  \int \frac{d^3q}{2q_0} \ f^{Bose}(q_0;T) \ 
{\rm Im} \Pi_{\rm em}^{I=1}(M,q;\mu_B,T)
\eeq
in the isovector channel, which amounts to a comparison
of  $\pi\pi$ annihilation (or $\rho$ decays) in the HG
versus $q \bar q$ annihilation in the QGP phase.
Starting from the free $\pi\pi$ and $q\bar q$ rates, which are obviously
very different from each other, the respective in-medium corrections lead 
to  (i) a characteristic low-mass enhancement in both the HG and QGP, 
as well as (ii) $\rho$-resonance melting (around $M=m_\rho$) 
and chiral mixing (1~GeV$\le M \le $~1.5~GeV) in the HG. 
As a result, the overall emissivities from the hadronic and Quark-Gluon 
phase look surprisingly similar around $T_c$, which has been 
interpreted as an in-medium reduction the quark-hadron duality
scale  $s_{dual}$~\cite{RW99}.     

\section{Dileptons at CERN-SPS Energies}
\subsection{Comparison to Data}
Under the assumption that the highly excited matter formed in 
central heavy-ion reactions reaches thermal equilibrium at some
typical formation time $\tau_0$ ($\sim$~1~fm/c at SPS), hydrodynamic
or thermal fireball simulations are the appropriate framework
for a consistent application of the thermal rates discussed in the 
previous section. Based on a fixed entropy per baryon as inferred 
from produced particle abundances, an isentropic thermodynamic 
trajectory $T(\mu_B)$ consistent with standard hadro-chemical 
freezeout analyses~\cite{pbm99,CR00} can be constructed including
a phase transition from the QGP to HG. 
An important ingredient in the fireball expansion
from chemical freezeout, $(T_{ch}, \mu_B^{ch})$ to the thermal one,
$(T_{th},\mu_B^{th})$, is the conservation of the observed particle 
multiplicities. This necessitates the build-up of finite 
meson-chemical potentials towards $T_{th}$, \eg,  
$\mu_\pi^{th}\simeq$~60-80~MeV at $E_{lab}=158$~AGeV.  
The contribution of thermal radiation to observed spectra
then takes the form
\beq
\frac{dN^{thermal}_{l^+l^-}}{dM}=\int\limits_{\tau_0}^{\tau_{fo}}
            d\tau \ V_{FB}(\tau) \int d^3q  \ \frac{M}{q_0} \  
                  \frac{dR_{l^+l^-}^{thermal}}{d^4q} \ Acc \ ,  
\eeq 
with $V_{FB}$ the fireball volume, and 
the factor $Acc$ accounts for experimental acceptance cuts.
In addition, contributions from long-lived hadron decays after 
freezeout have to included (the so-called 'cocktail'~\cite{na45-158}). 

\begin{figure}[!htbp]
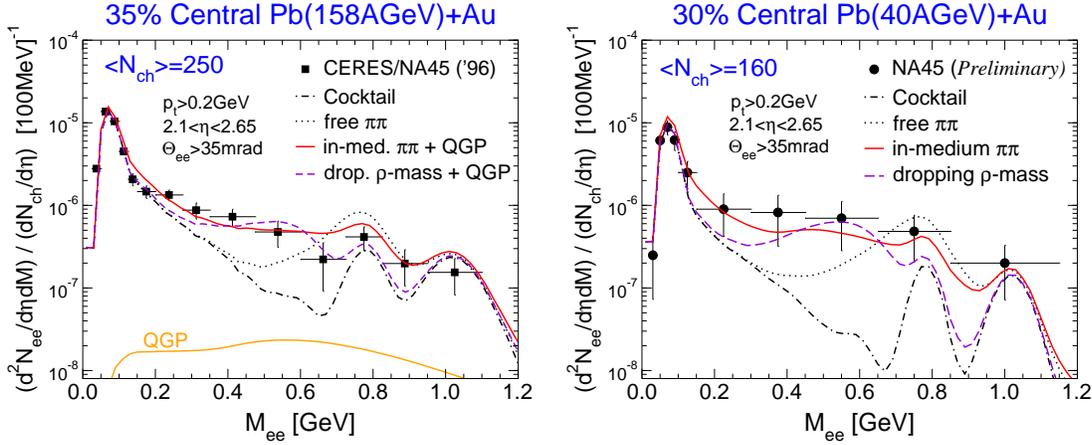

\vspace{0.1cm}
\bce
\epsfig{file=dlsQH69PbyT250-6.eps,width=7cm}
\hspace{0.2cm}
\epsfig{file=dlsQH69PbuT140.eps,width=7cm}
\ece
\caption{Thermal fireball calculations~\protect\cite{RW99} for (semi-) 
central Pb+Au at 158~AGeV and pertinent predictions for 40~AGeV compared 
to CERES/NA45 data~\protect\cite{na45-158,na45-40}.}
\label{fig_spsfb}
\end{figure}
An example of a thermal fireball calculation~\cite{RW99}
plus cocktail is shown in the left panel of Fig.~\ref{fig_spsfb}
and compared to CERES/NA45 data from Pb(158~AGeV)+Au collisions.
The enhancement over the cocktail cannot be described by adding
free $\pi\pi$ annihilation within the fireball. With the strong
medium effects as displayed in Fig.~\ref{fig_drho} ($\rho$-'melting')
a reasonable description is obtained. However, also the assumption
of a dropping $\rho$-mass reproduces the data. The QGP contribution
is small and insensitive to initial temperature and details
of the phase transition construction.
In the right panel of Fig.~\ref{fig_spsfb} the {\em predictions}
of the same approach~\cite{RW99} are found to be consistent with 
the measurements at lower SPS energies (40~AGeV) as well.  In fact, 
despite the lower pion multiplicities, the calculations imply 
a slightly larger signal in the $M\simeq 0.4$~GeV region due  
to larger medium effects induced by higher baryon densities in the 
hadronic phase. This is in line with the trend of the data.  

\begin{figure}[htbp]
\bce
\epsfig{file=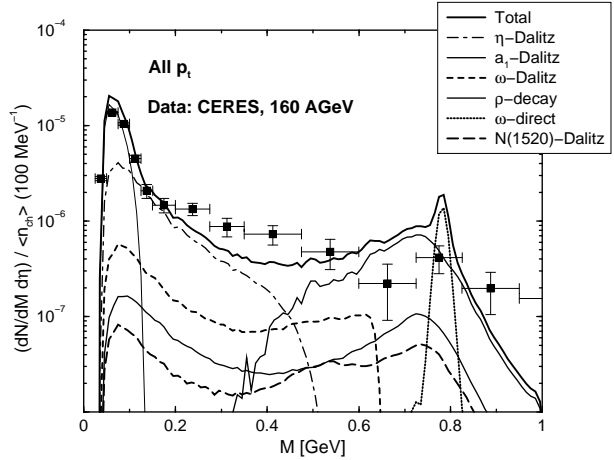,width=8cm}
\ece
\caption{Transport calculations~\protect\cite{Blei00} for Pb(40~AGeV)+Au;
the 158~AGeV data~\protect\cite{na45-158} have been included for 
orientation only.}
\label{fig_spstr}
\end{figure}
Another example of a low-mass dilepton prediction for the CERES
40~AGeV run is shown in Fig.~\ref{fig_spstr} in terms of a 
UrQMD transport calculation~\cite{Blei00}. Closer inspection reveals 
that the Dalitz decay contributions $\omega\to \pi^0 e^+e^-$ and 
$\eta\to \gamma e^+e^-$, which are prevalent around $M\simeq 0.4$~GeV,  
are a factor of 3-4 above the standard hadrochemical 
cocktail~\cite{na45-40}.  

\begin{figure}[htbp]
\bce
\epsfig{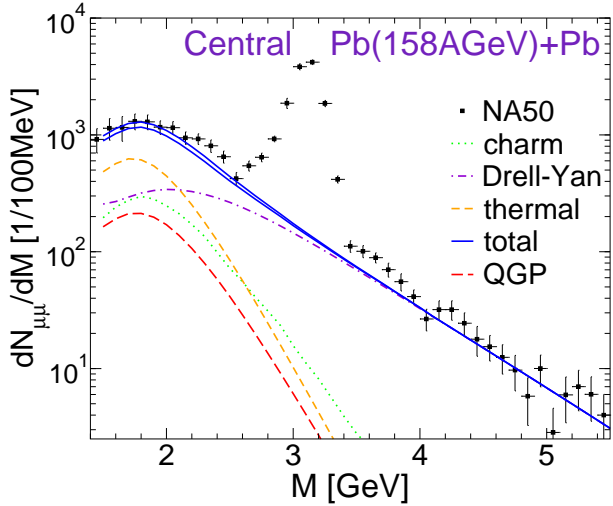}
\ece
\caption{Thermal fireball calculations~\protect\cite{RS00} 
compared to NA50 data~\protect\cite{na50-int}. Upper and lower total 
yields correspond to $T_0=195$~MeV and 220~MeV, respectively.}
\label{fig_spsint}
\end{figure}

Finally, we display in Fig.~\ref{fig_spsint} a thermal fireball 
calculation~\cite{RS00} for the intermediate-mass dimuon spectra 
measured by NA50 in central Pb(158~AGeV)+Pb. 
The 'dual' dilepton production rate based on eq.~(\ref{Piem}) has been
folded over the same space-time evolution (modulo centrality)
underlying the results of Fig.~\ref{fig_spsfb}\footnote{Note that
the experimental fact that hadron production in $e^+e^-$ annihilation
follows statistical (thermal) model predictions justifies the use of
the perturbative ('dual') rate in the intermediate mass region
even in the hadronic phases of a heavy-ion collision: the thermal
heat bath is the same state as produced in $e^+e^-\to hadrons$,
so that one can use time-reversal invariance to obtain the
$e^+e^-$ production rate.}.
As anticipated, the yield is much more sensitive the early
phases with a significant part of the yield emerging from a QGP
with preference for initial temperatures around $T_0\simeq 220$~MeV.

\subsection{Consistency of Models}
Let us first address the low-mass enhancement observed by CERES/NA45.
Under the premise that the relative importance of baryon density effects 
around midrapidity increases towards lower collision energies 
(larger baryon stopping and lower initial temperatures), the indication
for a large enhancement in the 40~AGeV data suggests that its origin lies 
in baryon-driven in-medium effects as predicted within the many-body
approach of refs.~\cite{RCW97,RW99} (unless dramatic nonperturbative 
effects occur in the QGP phase close to $T_c$, which seems unlikely
and finds no support in lattice results~\cite{Laer02}). At the same 
time, transport calculations predicted very little effect from higher 
baryon density~\cite{Blei00} and, in their present form, cannot
describe the 158~AGeV and 40~AGeV data simultaneously.   

Secondly, the NA50 dimuon enhancement at intermediate mass can be
explained by thermal radiation within the same (fireball) evolution
scenario~\cite{RS00} as the CERES data without additional parameters, 
albeit a with higher sensitivity to early (QGP) phases, as discussed 
above (see also refs.~\cite{GKP00,SSKG02}). In particular, an 
'anomalous' open-charm enhancement (over the expectation from 
$N$-$N$ collision scaling) is not required.   
In fact, also the $J/\Psi$ data of NA50~\cite{na50-jpsi} follow within 
a common framework~\cite{GR01} which combines suppression in a QGP 
and thermal production at $T_c$~\cite{pbm00}. 

Finally a remark on thermal photons is in order~\cite{PT02}.The WA98 
data~\cite{wa98} seem to require higher initial temperatures than the 
NA50 dimuons. However, photon production (both the hard component from 
$N$-$N$ collisions as well as the QGP rates) has a nontrivial
leading $\alpha_s$-dependence which implies relatively larger uncertainties
than in the dilepton sector. This equally applies to effects of (nuclear)
$k_t$-broadening (Cronin effect), which might be larger than assumed 
in current calculations and thus responsible for (part of) the 
discrepancy.

\subsection{Chiral Restoration?}
Let us briefly elaborate on some recent developments. 

In ref.~\cite{HY01} the so-called 'vector manifestation' of 
chiral symmetry restoration has been suggested: within the 
Hidden Local Symmetry framework, the chiral partner of the 
pion is identified with the longitudinal component of the 
$\rho$ meson. When applied within a one-loop renormalization 
group evolution towards the symmetry restoration point, 
the $\rho$ mass along with $f_\pi$ and $g_\rho$ go to zero.
In particular, VDM does not hold at finite $T$.

Rather different results are obtained in ref.~\cite{UBW01},  
where a linear $\sigma\pi\rho a_1$ model, consistently 
constructed to 1-loop order with realistic vacuum 
properties, has been applied at finite temperature. 
It is found that, even if the vacuum $\rho$ mass at tree level 
is entirely given through the $\langle\bar qq\rangle$ condensate, 
the vanishing of the latter at $T_c$ does not significantly
impact $m_\rho(T_c)$ due to different (leading) temperature 
dependencies of the two quantities~\cite{DEI90}.   
The main effect was rather a substantial broadening of both the 
$\rho$ and $a_1$ spectral functions. This is not inconsistent with
the chiral restoration scenario put forward in 
ref.~\cite{RW99}, where the degeneracy of vector and axailvector
correlators is realized through the in-medium reduction
of the quark-hadron duality threshold.   

\begin{figure}[htbp]
\bce
\epsfig{file=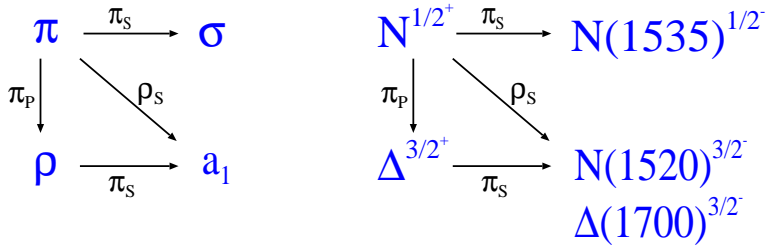,width=10cm}
\ece
\caption{Interaction scheme combining chiral and resonance excitations.}
\label{fig_schem}
\end{figure}

The relative importance of baryonic effects in 'melting'
the $\rho$ resonance calls for a better understanding of 
their relation to chiral symmetry. A possible scheme is
illustrated by the diagrams in Fig.~\ref{fig_schem}, which 
combines pionic $S$-wave excitations connecting chiral 
partners with pionic $P$-wave excitations representing
hadronic resonance physics.  
In particular, the analogy between the mesonic (left panel) and 
baryonic (right panel) sector identifies the prominent role
of the $\rho N$ $S$-wave resonances $N(1520)$ and $\Delta(1700)$.

\section{Predictions for RHIC}
The beginning of the collider era through the first
operation of BNL-RHIC has opened a new energy frontier 
in heavy-ion physics. First dilepton data are expected from the 
second run of Au+Au (completed in Nov.~2001) 
at the full energy of $\sqrt s =200$~AGeV.

\begin{figure}[htbp]
\bce
\epsfig{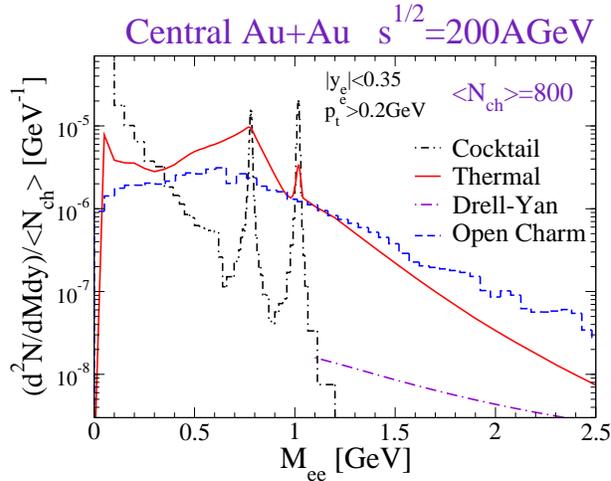}
\ece
\caption{Dilepton spectra from central Au+Au collisions at
full RHIC energy.}
\label{fig_rhic}
\end{figure}

Predictions for $e^+e^-$ spectra in central Au+Au are summarized in 
Fig.~\ref{fig_rhic} for the invariant-mass range from 0 to 2.5~GeV.
The thermal contribution~\cite{Ra01} has been evaluated within a 
fireball model with realistic values for the charged particle 
multiplicity~\cite{phob02} and an estimated $\bar p/p$-ratio of 
$\sim$75\%.  The radiation from the QGP phase (with initial temperature 
$T_0\simeq 380$~MeV) dominates the thermal yield at masses 
$M\gsim 1.5$~GeV. Its detectability will critically depend on the 
contribution from correlated open-charm decays, which in 
Fig.~\ref{fig_rhic} has been taken from a PHENIX event 
generator~\cite{Aver01} which is based on an extrapolation of 
$N$-$N$ collisions (using PYTHIA) at lower energies. If the spectrum 
of the $c$-quarks experiences significant softening (\eg, through 
energy loss in the QGP), the associated dilepton spectrum may 
be severely suppressed above $M=2$~GeV or so~\cite{Shu97}, opening
the window for the QGP signal.
In the low-mass region thermal radiation mostly originates 
from the hadronic phase and compares favorably to 
both open-charm and the hadronic decay cocktail after 
freezeout~\cite{Aver01} (here, subtraction of the combinatorial
background will be the main experimental problem).
In-medium modifications again 'melt' the $\rho$-resonance which
drives the enhancement below $M\simeq 0.6$~GeV~\footnote{Even with
the rather small {\em net} baryon densities at  midrapidities at 
RHIC, the copious abundance of baryon-antibaryon pairs contributes
significantly to the broadening~\cite{Ra01}.}. 
Similar effects lead to an in-medium signal from $\omega$-decays with 
an average width of $\sim$~50~MeV, almost seven times its vacuum value 
(less pronounced for the $\phi$).    

\section{Conclusions}
Driven by exciting data, the theoretical and phenomenological
analyses of dilepton production in high-energy reactions of heavy
nuclei have undergone continuous progress with increasing consensus
among most approaches. 
As for the low-mass region, the CERES/NA45 data at full SPS energy 
(158~AGeV) require strong medium
modifications of the $\rho$-meson indicating that one is indeed
probing strong interaction matter in the vicinity of $T_c$ (with 
small contributions from the QGP phase itself).  
The new 40~AGeV data support the importance of baryon-driven 
effects, while their relation to chiral symmetry (especially 
for resonances) remains to be better understood. '$\rho$-melting' 
with 'quark-hadron duality' towards  $T_c$ remains a viable scenario
of chiral restoration.    
At intermediate masses, the NA50 enhancement 
consistently emerges from the same thermal source, but with significantly 
larger sensitivity to the QGP phase pointing at initial temperatures
$T_0\ge 200$~MeV. 
New exciting insights can be expected soon from 
PHENIX at RHIC, HADES at SIS and farther into the future from 
ALICE at LHC. This raises the hope of a systematic understanding of 
dilepton emissivities across the QCD phase diagram.

\vspace{0.5cm}

\noindent
{\bf Acknowledgment} \\
I would like to thank the organizers for a very interesting
and stimulating conference. This work has been supported by 
the US-DOE grant DE-FG0288ER40388.


\end{document}